\documentclass[runningheads,a4paper]{llncs}

\usepackage{amssymb}
\usepackage{graphicx}

\usepackage{url}
\urldef{\mailsa}\path| mahyunst@yahoo.com, samn@ftsm.ukm.my |
\newcommand{\keywords}[1]{\par\addvspace\baselineskip
\noindent\keywordname\enspace\ignorespaces#1}

\begin{document}

\mainmatter  

\title{A Methodology to Extract Social Network from the Web Snippet}

\titlerunning{A Methodology to Extract Social Network from the Web Snippet}

%
%
\author{Mahyuddin K. M. Nasution$^{1,2}$%
\thanks{}%
\and Shahrul Azman Noah$^{2}$}
\authorrunning{M. K. M. Nasution and S. A. Noah}

\institute{$^{1}$Departemen Teknologi Informasi, FASILKOM-TI,\\
Universitas Sumatera Utara, Padang Bulan 20155 USU, Medan, Indonesia.\\
$^{2}$Knowledge Technology Research Group, CAIT-FTSM\\
Universiti Kebangsaan Malaysia, Bangi 43600 UKM Selangor Malaysia.\\
\mailsa\\}

%
%

\toctitle{draf}
\tocauthor{}
\maketitle

\begin{abstract}
The Web has been chosen as a basic infrastructure to gain the social structure information, through the social network extraction, from all over the world. However, most of the web documents are unstructured and lack of semantics. Moreover, that network is subject to all kinds of changes and dynamics, and a network can be very complex due to the large number of nodes and links Web contains. In this paper, we discuss a methodology that meant to assists in extracting and modeling the social network from Web snippet. As the manual social network extraction of web documents is impractical and unscalable, and fully automated extraction are still at the very early stage to be implemented, we proposed a (semi)-automatic extraction based on the superficial methods.
\end{abstract}

\section{Introduction}

Social network is a model for representing relations between individuals, groups or organization \cite{nasution2010}. Today, the study of social networks has received as much attention as studies of computer network or computer mediated communication: When a computer network connects people or organizations, it is a social network \cite{wellman1999} and the use of Web therefore is steadily gaining ground in the study of social networks. Web contains an enormous amount of information, has not been only becoming the largest database in computer network ever seen in the history, but is becoming more and more complex while its size continuous to grow at a remarkable rate, for which social network extraction algorithms are being developed on a large scale \cite{mika2007}. Social network extraction is an establishment of a technology to identify and describe special content of web pages: actors and their relations \cite{nasution2012}. That is representation of web pages and query that are enriched with semantic information of their respective contents. Therefore, extracting social network from Web depends heavily on the cooccurrence whereby search engines play a role of a front gate to access the Web, whereby a resultant social network can be modeled very naturally by a graph $G(V,E)$, i.e. a set of nodes $V\ne\emptyset$ is a representation of actors and edges $E$ is to represent the relations. However, in statistical literature \cite{robins1999}, social network model assumes that there are $n$ actors and innformation about binary relation between them in $n\times n$ matrix $M$, where $e_{ij}= m_{ij} \in M$ is $1$ $\forall e_{ij}\in E$ if $v_i, v_j\in V$ are adjacent, $0$ otherwise. If matrix $M$ is asymmetric, then the relations are represented as directed edge, $e_{ij}\ne e_{ji}$. Therefore, the social networks grows with $O(n^2)$. Moreover, each search engine limited the number of queries to $m$ per day per IP (Internet Protocol) address. In this paper therefore we present a methodology for extracting social network from Web. Social network extraction with considering the scalability of computational complexity and submitting the queries.

\section{Related Work}
The notion of social network extraction concept is not completely clear in some literature \cite{kautz1997,diesner2005}, but the network concept is one of the well-defined paradigms of modern era for understanding the phenomena in world that is a concept based on graph theory clearly wehreby for building a social network simply consists of two phases: after fixed nodes, edges added between them. In Artificial Intelligence (AI), two research streams for extracting social network from various sources of information exists: supervised \cite{diesner2005,mccallum2005} and unsupervised\cite{kautz1997,matsuo2007}. The supervised methods utilize the Cartesian product for clustering the actors $A$ into the network. The concept of clustering is $\gamma : A\times A \rightarrow R$ such that $\gamma(a_i,a_j) \in R$, and $a_i,a_j\in A$. $A$ is a set of actors, and $R$ is a set of relations. However, the clustering approach is difficult to generate labels of relations in a network. The supervised methods employ a function $\lambda$ for classifying $Z$, i.e. $\lambda : Z\rightarrow C$ such that $\lambda(z) = z$, $z\in Z$, and $c\in C$ is a class label, where $C = \{c_1,c_2,\dots,c_{|C|}\}$ is a data set as special target attributes, $|C|\ge 2$ is the number of classes, $Z\cap C=\emptyset$ and $Z$ is a set of descriptions of actors in $A$. The classification approaches however only concern with extraction of network based on predefined labels only, and thus cannot be adapted to the other descriptions of relations.

In the present study, we generally refer to social network extraction as follows.

\begin{definition}
The social network extraction is a mapping $\gamma_e = \gamma_1\gamma_2$ for acquiring the social networks, i.e. $SN = \langle V,E,A,R,Z,\gamma_1,\gamma_2\rangle$ that satisfies the following conditions
\begin{enumerate}
\item $\gamma_1 = A\rightarrow V$, $v=\gamma_1(a)$ $\forall a\in A$ $\exists! v\in V$.
\item $\gamma_2 = R\rightarrow R$ so that $e_j = \gamma_2(r_k(a_1,a_2)) = \gamma_2(Z_{a_1}\cap Z_{a_2})$, $e_j\in E$, $r_k\in R$ $\forall a_1,a_2\in A$, $Z_{a_1}, Z_{a_2}$ are subsets of $Z$ and is a set of attributes of $a_1$ and $a_2$ respectively.
\end{enumerate}
\end{definition} 

This definition ranges from contents relations to strength relations. The content can be represented as the resources that is exchanged, directed or undirected relations is to represent all expressions of interactions in symmetrical or asymmetrical, and strength relations is that respect to event frequencies of each pair of actors \cite{nasution2011b}. Some social network systems such as ReferralWeb \cite{kautz1997}, Polyphonet \cite{matsuo2007}, etc. have been developed to extract the social networks based on the similarity measures. In general the similarity measure considers three statistical values of submitted query (hit count): singleton $|{\bf a}|$ and doubleton $|{\bf a}\cap{\bf b}|$. A pseudocode that measures the strength relation ($sr$) of two persons based on singleton and doubleton \cite{kautz1997,matsuo2007} is\\

\noindent
Function $sr(a,b)$:\\
$|{\bf a}|\leftarrow query(a)$\\
$|{\bf b}|\leftarrow query(b)$\\
$|{\bf a}\cap{\bf b}|\leftarrow query(a,b)$\\
return $sim(|{\bf a}|,|{\bf b}|,|{\bf a}\cap{\bf b}|)$.\\

Therefore, a network with $n$ actors requires $3n^2$ queries, but because of strength relations is symmetric (undirected edges) the network grows with $O(nm)$, $n>m$. However, the number of queries remains a great problem for limitation of search engines.

Definition 1 also explicitly explains that the social network extraction needs a strategy for disambiguating names. Name disambiguation problem has relationships with semantic because of (a) different actors can share the same name (lexical ambiguity), and (b) a single entity can be designated by multiple names (referential ambiguity) \cite{yang2007}. The ambiguity related to natural language texts and the unreliability of individual observations taken from an uncontrolled respository such as web pages, it generally limits performance of any extraction methods. One of solutions for both tasks of name disambiguation in unsupervised stream was by adding keywords $kw$ into query $q$ \cite{jin2007}, i.e.
\[
q(a) = a,
\]
\[
q(a,kw_1) = a+kw_1,
\]
or
\[
q(a,kw_1,kw_2) = a+kw_1+kw_2.
\]
Therefore, a query that consists of two actor names and $q = a+kw_1$ take same behavior as like as a doubleton \cite{bekkerman2005}. It means that if there are two actor names in query, one name be the keyword to another. Formally, we can define it as $|{\bf a}\cap{\bf b}|\leftarrow q(a,b) = a+b$ $\forall a,b\in A$. The candidate keywords can be categorized into stable and flexible attributes (descriptions). Email address: mahyuddin@usu.ac.id for example is a stable attribute, while "extract", "teluk", "noah", etc. are the flexible attributes of named actor "Mahyuddin K. M. Nasution". Some attributes can be extracted from web pages or web snippets. We developed a micro-cluster approach to generate some keywords from web snippets as name disambiguation for extracting social networks from Web \cite{nasution2010}. One of approaches to resolve the name ambiguity in social networks extraction is by using well-defined actor name as a seed, where we developed a superficial method in unsupervised stream for extractiong social network that can be employed with developing a general pattern based on association rule \cite{nasution2011a}. This enhanced superficial method not only generates a network, but provides labels of networks. In this case of labeling network, there are some researchers developed an approach that not only assigns strength of its relation, but assigns underlying relations exists behind link where the extracted labels from web snippets \cite{mori2006} and from URLs address \cite{nasution2010}. A note for applying a rule in pattern that is the heuristic rule based method requires the predefined rule for each specific type of social network extraction problem, which is not adaptive for different situations. Therefore, in line with the superficial method we develop an approach for dealing with the unstructured and lack of semantics in Web, and derive the rules based on context general for reducing the scale of computation complexity to be $O(n)$.

\section{Social Network Extracting Methodology}
In this section first we present the bases of our methodology. The Web is the information source. Many classical social neworks extraction have performance limitations due to the use of a typically reduced corpus \cite{kautz1997,diesner2005}. The use of massive amounts of heterogeneous data can bring benefits to unsupervised methods development and can minimize the constraints regarding the availability of information for any domain analyzed. This idea is supported by current social studies in which it is argued that collective knowledge is much more powerful than individual knowledge. The Web is the largest repository of information available that is build upon the global network which connects all people around the world together, where most data on the Web are so unstructured and they can only be understood by humans, but the amount of data is so huge and they can only be processed efficiency by machines. Thus, a direct analysis of such an enormous repository is impracticable and unscalable, and fully automated tools are still at the very early stage to be implemented. Therefore, web search engines can be exploited as effective Web information retrieval and extraction tools. On Web the search engines are extensively important to help users to find relevant information whereby the query has become the leading paradigms to find the information. Based on content analysis, there are three features of web pages that are considered as the basis of extracting social relations \cite{nasution2010,mika2007}: (a) Co-occurrences of names of actors in web pages are an evidence of relationships and are a more frequent phenomenon. For example, author co-authorship in papers. (b) The linking structure of the Web is a proxy for real world relationships as links are chosen by the web page author and connect to other information sources that are considered authoritative and relevant enough to be mentioned. It looks like citation of a paper. (c) URL address indicates the layered structure of a Web site which can be logically shown as a hierarchy, wehre web site editors usually tend to put similar or related web pages as close as possible underlying relations among entities such as a cooccurrence.

Some features are related with Web snippet (snippet). $S$ is a snippet, $S$ represents a web page $\omega$ then $S$ contains at least three elements: (a) A composition of URL contains a set of tokens $U = \{s,d_1,\dots,d_m,p_1,\dots,p_{n-1}\}$ satisfying a structure $s://d_m.\cdots.d_2.d_1/p_1/\cdots/p_{n-1}$ is a string consists of scheme, authority, and path. (b) A title of web page or $tit$, i.e. a very important piece of information in terms of providing a meaningful summary of the page for the search engines, which display title content in the search results. (c) An abstract or $abs$ contains apart of web page body, i.e. everything that relate to first paragraphs.

Secondly, a proposed (semi)-automatic methodology for extracting social networks from the Web. Three stages have defined: (i) the detection of relations, differentiating actors relations in $A\times A$ for $L_{ab}\ne\emptyset$ and hit counts in $H_{ab}$, (ii) the keywords extraction and selection of attributes values for each actor: $L_a$, $H_a$, $K_w$ are a list of snippets, hit count, and list of keywords, respecively based on our previous work \cite{nasution2010} and this do not explain here, and (iii) the identification of measurement units: Strength relation extraction. The latest stage, aimed to compare some superficial methods, are only considred for scale $< n(n-1)/2$ or no add new query.

\subsection{Relation detection and Underlying relation}
We use the concept of cooccurrence as the underlying relation for enhancing the superficial method and here we generate a rule based on an associatio rule \cite{bekkerman2005}. In text literal, $Z$ contains a query of two actor names $q(a,b)$ for all $a,b\in A$, a doubleton ${\bf a}\cap{\bf b}$, and a snippet $S$, where $X = \{q(a,b)\}$ and $Y = \{|{\bf a}\cap{\bf b})>0,t_a,t_b\in S\}$. We have a rule as follows,\\

\noindent
{\bf Rule 1.}
$((q(a,b)\Rightarrow |{\bf a}\cap{\bf b}|>0)\Rightarrow t_a,t_b\in S)$\\

Rule 1 detects the existence of relations among two actors ($a,b\in A$), if $|{\bf a}\cap{\bf b}|>0$ and search terms $t_a$ and $t_b$ in $S$. In other words, the implication $((q(a,b)\Rightarrow |{\bf a}\cap{\bf b}|>0)\Rightarrow t_a,t_b\in S)$ is TRUE if $(q(a,b)\Rightarrow |{\bf a}\cap{\bf b}|>0)$ is also TRUE. Applying this rule to a list $A$ containing $n$ names will detect $n(n-1)/2$ potential relations, and it involves $n(n-1)/2$ queries. In this case, in constructed queries the search term of names are written between double quotes ("...") in order to force the search engine to provie exact pattern matches. Thus, the queries was reduced until $n(n-1)/2$ times of submitting to any search engine. Each query is executed and, as a result, a list of web snippets to be analyzed is retrieved. On the same occasion, this approach provides an opportunity for gathering clues of relationship between two actors. By simple approach we can extract the labels of networks that are useful for describing relations in networks as surrounding current context in which actors of interest cooccur on the Web. Therefore, Rule 1 we use also to collect the snippets into a list $L$ as clues of relation, i.e. a list of doubleton snippets $L_{ab}$, as the underlying relations among actors in pair ($a,b \in A$), is a collection of web snippets containing $t_a$ and $t_b$ for each snippet. $L_{ab} = \{S_i|i=1,\dots,m$; $t_a,t_b\in S_i$ $\forall a,b \in A\}$. This expresses that if the relations between actors in pair exists then $L_{ab}\ne\emptyset$.

\subsection{Strength relation and underlying strength relations extractions}
The concept of strength relations generally is by exploiting the similarities of actor pairs. The vector in a space of event gives some means to demarcate social network whereby the estimation has a role depends on a certain threshold. Therefore, depending on the measurement, the resultant social network varies. In this study, we can use some similarity measures. Thus a relatin that exploits two singletons and a doubleton, i.e. ${\bf a}$, ${\bf b}$, ${\bf a}\cap{\bf b}$ are subsets of $\Omega$, by using a similarity measure $sim(t_a,t_b)$ in $[0,1]$, we call a strength relation $sr$, if it satisfies the conditions: (1) $|{\bf a}\cap{\bf b}|\le|{\bf a}|$ and (2) $|{\bf a}\cap{\bf b}\le|{\bf b}|$. Using Jaccard coefficient for example we obtain $sim(t_a,t_b)=|{\bf a}\cap{\bf b}|/(|{\bf a}|+|{\bf b}|-|{\bf a}\cap{\bf b}|)$, $(|{\bf a}|+|{\bf b}|-|{\bf a}\cap{\bf b}|)>0$. In this case, $|{\bf a}|\leftarrow q(a)$, $|{\bf b}|\leftarrow q(b)$ and $|{\bf a}\cap {\bf b}|\leftarrow q(a,b)$. However, the hits show always the bias of cooccurrence. Thus, for estimating the strength of their relation by cooccurrence of their two names, we add the keyword for each actor, i.e. $|{\bf a}\cap{\bf kw}_1|\leftarrow q(a,kw_1)$, $|{\bf b}\cap{\bf kw}_2|\leftarrow q(b,kw_2)$, and $|{\bf a}\cap{\bf kw}_1\cap{\bf b}\cap{\bf kw}_2|\leftarrow q(a,kw_1,b,kw_2)$. We call it as Strength Relations with keyword (SRwK). Other approach for identifying strength relations among actors is to exploit the URL address we call it as Underlying Strength Relation (USRs)\cite{nasution2010}. URL of web pages which provides and indicator of its logical position in the hierarchy structure that can be considered as the underlying strength of the relationship and this proposed approach migh generate multiple labels for edge from tokens of URL or DNS (domain name system).

To attain the scalability of some methods for extracting strength relations above, we allow Rule 1 using search engine and reduce the complexity of SR, SRwK or USR methods. That is less than or equal to $mn$ potential relations whereby number of queries also is $<3nm$. Rule 1 enables us to investigate the relations of a larger set of actors where some queries for generating the relations has been eliminated.

\section{Conclusion and Future Work}

This article presents a practical methodology for extracting social network based on superficial methods (in unsupervised research strem), quality measures for clustering the actors and relations between them with keywords, singleton and doubleton, and URLs address, we obtain the scale $\le n(n-1)/2$. The social network extraction process relies on a dynamic knowledge in Web, thus this is still under development, and research work must be completed and enriched. In particular, a careful next study of measures for their own and compared to each other of superficial methods, mainly to consider complexity and number of submitted queries.

\end{document}